\documentclass{kluwer}    % Specifies the document style.
\usepackage{graphicx}
\usepackage{lineno}

\newcommand{\next}{\indent\newline\indent\newline}

\def\EPSFIG[#1]#2#3#4{		%
\begin{figure}[H]		%
\begin{center}			%
\includegraphics[#1]{#2}	%
\end{center}			%
\caption{#3}			%
\label{#4}			%
\end{figure}			%
}				%
\newdisplay{guess}{Conjecture}

\begin{document}                                                                                   
\begin{article}
\begin{opening}         
\title{A New Sensitivity Analysis and Solution Method for Scintillometer Measurements of Area-Averaged Turbulent Fluxes} 
\author{Matthew \surname{Gruber}\email{matthewgruber@gi.alaska.edu}}
\author{Gilberto J. \surname{Fochesatto}\email{foch@gi.alaska.edu}}
\institute{Department of Atmospheric Science, College of Natural Sciences and Mathematics, Geophysical Institute, University of Alaska Fairbanks}

\begin{ao}
903 Koyukuk Dr.
99775, Fairbanks, AK.
Ph: 1-907-474-7602
Fx: 1-907-474-7290
\end{ao}

\begin{motto}
``Iteration, like friction, is likely to generate heat instead of progress.'' - George Eliot
\end{motto}

%\runningauthor{Gruber and Fochessatto}
%\runningtitle{A Cure for Iteration}
%\institute{Geophysical Institute}
%\date{June 10, 2012}

\begin{abstract}
Scintillometer measurements of the turbulence inner-scale length $l_o$ and refractive index structure function $C_n^2$ allow for the retrieval 
of large-scale area-averaged turbulent fluxes in the atmospheric surface layer. 
This retrieval involves the solution of the non-linear set of equations defined by the Monin-Obukhov similarity hypothesis.   
A new method that uses an analytic solution to the set of equations is presented,
which leads to a stable and efficient numerical method of computation that has
the potential of eliminating computational error.
Mathematical expressions are derived that map out the sensitivity of the turbulent flux measurements to uncertainties
in source measurements such as $l_o$. These sensitivity functions differ from results in the previous literature; the reasons for the differences
are explored.
\end{abstract}


\end{opening}           
\linenumbers
\section{Introduction}

Scintillometers detect fluctuations in the intensity of a beam of light that passes through a path length of 
50 m to 5000 m of near-ground turbulence in the surface layer \cite{KLEISSL2008}. These fluctuations are related to
the structure function of the index of refraction $C_n^2$, and the turbulence inner-scale length
$l_o$ \cite{TATARSKI,HILL1988L0,SASIELA}. The index of refraction is a function of temperature and humidity; thus $C_n^2$
can be decomposed into structure functions of temperature $T$ and humidity $q$ as $C_T^2$, $C_{Tq}$ and $C_q^2$. Scintillometer 
wavelengths are selected that are each more sensitive to fluctuations in one variable (such as temperature) than others (such as 
humidity), so that $C_T^2$, $C_{Tq}$ and $C_q^2$ may be resolved.
For example, intensity fluctuations of visible and near-infrared beams are more sensitive to temperature fluctuations than humidity
fluctuations, while microwave beams are more sensitive to humidity fluctuations \cite{ANDREAS3}. 
Structure functions such as $C_n^2$ are described in 
\inlinecite{TATARSKI}, and represent the strength and spacial frequency of perturbations in variables; 
thus $C_n^2$ is a measure of turbulence intensity weighted by the susceptibility of the index of refraction of the medium to changes in variables such as temperature and humidity.\newline

The goal of this study is to solve for the sensible heat flux $H_S$ and the momentum flux $\tau$ as functions
of source measurements such as $C_n^2$ and $l_o$, as well as to quantify
the propagation of uncertainty from source measurements to the calculated values of 
$H_S$ and $\tau$. Another type of turbulent flux is the latent heat flux $H_L$. 
The turbulent fluxes are given by

\begin{eqnarray}
H_S & = & -\rho c_p u_\star T_\star 			,	\label{SENSIBLEHEAT} \\
H_L & = & - L_v u_\star q_\star 			,	\label{LATENTHEAT} \\ 
\tau & = & \rho {u_\star}^2 				,	\label{MOMENTUM}
\end{eqnarray}
where $T_\star$ and $q_\star$ are the temperature and humidity scales, $u_\star$ is the friction velocity, 
$\rho$ is the density of the air, $c_p$ is the specific heat at constant pressure, and $L_v$ is the latent heat of vaporization. 
Determining area-averaged turbulent fluxes involves solving for $T_\star$ and $q_\star$, which are related to the path-length scale 
structure-function measurements through the non-linearly coupled 
Monin-Obukhov similarity equations \cite{SORBJAN}. This procedure also involves solving for 
$u_\star$ in Eqs. \ref{SENSIBLEHEAT}, \ref{LATENTHEAT} and \ref{MOMENTUM}. The friction velocity 
$u_\star$ can be related either to path-length scale $l_o$ measurements 
as with displaced-beam scintillometer strategies described in \inlinecite{ANDREAS1992}, 
or to the wind profile and roughness length with large-aperture scintillometer strategies via the Businger-Dyer relation 
\cite{PANOFSKY,SORBJAN,LAGOUARDE,HARTOGENSIS2003}.\newline

We consider here a displaced-beam scintillometer strategy in which path-averaged measurements of $C_n^2$ and $l_o$
are obtained. Other required measurements 
include temporally-averaged pressure $p$, temperature $T$, humidity $q$, as well as the 
height of the beam above the underlying terrain $z$.
Thus $C_n^2$, $l_o$, $p$, $T$, $q$ and $z$ are referred to as the 
source measurements. Each of these measurements demonstrates temporal and spacial variability as well as
measurement uncertainty. Uncertainty
propagates from the source measurements to the derived variables via the 
set of equations being considered.
Uncertainties in $l_o$ and $C_n^2$ are described in \inlinecite{HILL1988L0}, while uncertainties in $p$, $T$ and $q$ depend on the particular instrument being used. 
Here, we explore the use of scintillometers over flat and homogeneous terrain, thus the height of the beam $z$ is considered to be a single value with its associated uncertainty.
While $C_n^2$ and $l_o$ are representative of 
turbulent fluctuations along the whole beam, $p$, $T$ and $q$ are typically point measurements representative of localized areas near 
their respective instruments.\newline

Applications for scintillometers 
include agricultural scientific studies such as \inlinecite{HOEDJESOLIVES} and \inlinecite{FOKEN}, and aggregation of surface measurements to satellite-retrieval scales for weather
prediction and climate monitoring as in \inlinecite{BEYRICH2002} and in \inlinecite{MARX}.  
The unique spacial scale of scintillometer measurements gives them the potential for a key role in bridging the gap between 
ground-based instruments with footprints on the order of $100$ $\mbox{m}^2$ and model and satellite-retrieval scales on the
order of $1$ $\mbox{km}^2$.\newline

The scale of scintillometer measurements introduces an additional complexity in the retrieval of the turbulent fluxes.  
This retrieval combines the large-scale scintillometer measured variables $C_n^2$ and $l_o$ with source measurements that are not necessarily 
representative of the same scale. The only exception to be considered is the atmospheric pressure $p$. 
In particular, measurements of $T$ and $q$ may be representative of smaller footprints around their respective instruments. 
Specifically, assuming that variables such as average temperature T represent the
entire beam path introduces a form of uncertainty.
This uncertainty is somewhat similar to a systematic error, although it may be difficult to 
quantify because of its temporal variability.\newline

Of previous scintillometer sensitivity studies, some stand out as possibly contradicting each other.
For instance, the conclusion of the error analysis in \inlinecite{MORONI} for a $l_o$ and $C_n^2$ strategy was that
\textit{``The Monte Carlo analysis of the propagation of the statistical errors shows that there is only moderate sensitivity of the flux calculations to the initial errors in the measured quantities.''}
The error analysis of \inlinecite{ANDREAS1992}, however, results in sensitivity functions that feature singularities. 
The sensitivity functions presented there
imply that the resolution of $u_\star$ and consequently of $H_S$, $H_L$ and $\tau$ by scintillometer
$l_o$ and $C_n^2$ measurements 
is intrinsically restricted to low precision over a certain range of environmental conditions.
While these two studies use different methods and present results over slightly different ranges in variables, 
they produce sensitivity functions that for the same range differ significantly.\newline

In Sect. \ref{body} below, we decouple the set of equations including those of the Monin-Obukhov similarity hypothesis
for $l_o$ and $C_n^2$ scintillometer strategies for the example of unstable surface-layer conditions
to arrive at single equations in single unknowns. 
The variable inter-dependency is mapped out as illustrated by tree diagrams.
In Sect. \ref{RESULTS}, we take advantage of the mapped out variable inter-dependency to guide us in using the chain rule to solve the global 
partial derivatives in sensitivity functions to investigate error propagation. 
We produce sensitivity functions for $H_S$, $\tau$ and $u_\star$ as functions of both $l_o$ and $z$. In Sect. \ref{DISCUSSION} we explore
the ramifications of our results and compare them to previous literature, and we give conclusions in Sect. \ref{CONCLUSIONS}.

\section{Measurement Strategy Case Study: Displaced-Beam Scintillometer System in Unstable Conditions}\label{body}

We consider here a two-wavelength system as introduced in \inlinecite{ANDREAS1989}, where one of the scintillometers measures both $l_o$ and $C_n^2$ as in \inlinecite{ANDREAS1992}.
With this strategy, our measurements can resolve humidity and temperature fluctuations separately since 
the two scintillometers have different wavelengths $\lambda_1$ and $\lambda_2$ that have differing sensitivities in the index of refraction
to humidity and temperature. This technique therefore requires fewer assumptions than the
corresponding single-wavelength strategies as seen in \inlinecite{ANDREAS1989}.\newline

The following set of equations
determines $T_\star$, $q_\star$ and $u_\star$ from the source measurements, and subsequently determines the turbulent fluxes:

\begin{eqnarray}
\rho & = & \frac{p}{RT}											,		\label{IDEALGAS} \\
l_o & = & \frac{(9\Gamma(1/3)K D(\rho,T))^{3/4}}{\epsilon^{1/4}}					,		\label{LOEPSILON} \\
\zeta & = & \frac{zg\kappa}{{u_\star}^2T}\left(T_\star+\frac{0.61T}{\rho+0.61q}q_\star\right) 	,		\label{lmo} \\
{u_\star}^3 & = & \frac{\kappa z \epsilon}{\phi(\zeta)} 						,		\label{ustar} \\
{C_{n_1}^2} & = & z^{-2/3}g(\zeta)(A_1(\lambda_1,p,T,q)T_\star+B_1(\lambda_1,p,T,q)q_\star)^2 	,		\label{cn1} \\
{C_{n_2}^2} & = & z^{-2/3}g(\zeta)(A_2(\lambda_2,p,T,q)T_\star+B_2(\lambda_2,p,T,q)q_\star)^2 	,		\label{cn2}
\end{eqnarray}
where $g$ is the local acceleration due to gravity, $\Gamma$ is the Gamma function, $\epsilon$ is the turbulent energy dissipation rate, 
$R$ is the specific gas constant, $\kappa$ is the von K\'arm\'an constant, $\zeta\equiv z/L$, where $L$ is the Obukhov length,
$K$ is the Obukhov-Corrsin constant, $\nu(T,\rho)$ is the viscosity of air and $D(T,\rho)$ is the thermal diffusivity of air 
(Andreas, 1989; 1992; 2012)
%\cite{ANDREAS1989,ANDREAS1992,ANDREAS2012}. 
$C_{n_1}^2$ and $C_{n_2}^2$ are structure functions of the refractive index for the separate wavelengths $\lambda_1$ and $\lambda_2$.
Eqs. \ref{IDEALGAS} and \ref{LOEPSILON} determine $\epsilon$ directly from $l_o$ and the other source measurements.
Inherent in Eqs. \ref{cn1} and \ref{cn2} is the assumption that $C_{Tq} = \sqrt{{C_T^2}{C_q^2}}$, which is validated previously \cite{HILL1989,ANDREAS3}.\newline

The similarity functions $g(\zeta)$ and $\phi(\zeta)$ are given by

\begin{eqnarray}
g(\zeta) = a(1-b\zeta)^{-2/3}			,	\\ 
\phi(\zeta) = (1+d(-\zeta)^{2/3})^{3/2} 	,
\end{eqnarray}
for $L<0$ 
which corresponds to unstable conditions. The form of the similarity functions and their parameters follow from \inlinecite{WYNGARD1971} and \inlinecite{WYNGARD1971phi}; the values are 
taken to be $a=4.9$, $b=6.1$, and $d=0.46$ \cite{ANDREAS1988}.\newline
%, and by $g(\zeta) = a(1+c(\zeta)^{2/3})$ and $\phi(\zeta)=(1+h(\zeta)^{3/5})^{3/2}$ for $l>0$ which corresponds to stable conditions. 

The source measurements may not 
determine the sign of $L$, which is unknown a priori for every set of source measurements at any one time interval.
We follow \inlinecite{ANDREAS1989} in solving for $T_\star$ and $q_\star$ from Eqs. 
\ref{cn1} and \ref{cn2}, making sure to note that the signs of $\left(A_{1,2}T_\star+B_{1,2}q_\star\right)$ are not yet solved by introducing
unknowns $\mbox{sign}_1$ and $\mbox{sign}_2$:

\begin{eqnarray}
\frac{\mbox{sign}_1\sqrt{{C_{n_1}^2}}z^{1/3}(1-b\zeta)^{1/3}}{\sqrt{a}} & = & A_1T_\star\left(1+\frac{B_1}{A_1}\frac{q_\star}{T_\star}\right) 	, \\ 
\frac{\mbox{sign}_2\sqrt{{C_{n_2}^2}}z^{1/3}(1-b\zeta)^{1/3}}{\sqrt{a}} & = & A_2T_\star\left(1+\frac{B_2}{A_2}\frac{q_\star}{T_\star}\right)  	,
\end{eqnarray}
where the roots on the left-hand side are considered to be positive. Following \inlinecite{ANDREAS1989}, these can be re-arranged to isolate $T_\star$ and $q_\star$ with the as yet undetermined signs:

\begin{eqnarray}
T_\star & = & \frac{(1-b\zeta)^{1/3}z^{1/3}}{\sqrt{a}}\left(\frac{\mbox{sign}_1\sqrt{{C_{n_1}^2}}B_2-\mbox{sign}_2\sqrt{{C_{n_2}^2}}B_1}{A_1B_2-A_2B_1}\right) \label{tstar} 	,\\
q_\star & = & \frac{(1-b\zeta)^{1/3}z^{1/3}}{\sqrt{a}}\left(\frac{\mbox{sign}_2\sqrt{C_{n_2}^2}A_1-\mbox{sign}_1\sqrt{{C_{n_1}^2}}A_2}{A_1B_2-A_2B_1}\right) \label{qstar}		,
\end{eqnarray}
where 

\begin{equation}
\mbox{sign}_{1,2} = \mbox{sign}[A_{1,2}T_\star(1+\frac{B_{1,2}}{A_{1,2}}\frac{q_\star}{T_\star})]		.
\end{equation}

It is useful to include the definition of the Bowen ratio as 

\begin{equation}
\beta\equiv H_S/H_L = \frac{\rho c_p}{L_v}\frac{T_\star}{q_\star}		.
\end{equation}
We can solve for $\beta$ as

\begin{equation}
\beta = E\left(\frac{\mbox{sign}_1\sqrt{C_{n_1}^2}B_2-\mbox{sign}_2\sqrt{{C_{n_2}^2}}B_1}{\mbox{sign}_2\sqrt{C_{n_2}^2}A_1-\mbox{sign}_1\sqrt{C_{n_1}^2}A_2}\right)		,
\end{equation}
where $E(T,p)=\rho c_p/L_v$. It is useful to consider $\beta$ as well as $\zeta$  as unit-less independent variables in our sensitivity analyses that represent certain meteorological 
regimes. They represent the ratio of the
sensible to latent heat fluxes and an indicator of surface-layer stability, respectively.\newline

Since we are considering unstable conditions, we have $\zeta<0$ since $L<0$, so from Eq. \ref{lmo} we have 

\begin{eqnarray}
T_\star(1+\frac{0.61T}{\rho+0.61q}\frac{q_\star}{T_\star})<0 			,	\\
(1-b\zeta)>0									,	\\
(1+d(-\zeta)^{2/3})^{3/2}>0							,
\end{eqnarray}

We begin decoupling the set of equations by taking Eqs. \ref{tstar} and \ref{qstar} 
and substituting into Eq. \ref{lmo}, then cubing the resulting equation as well as squaring Eq. \ref{ustar} to arrive at

\begin{eqnarray}
\zeta^3 & = & \frac{z^4g^3\kappa^3(1-b\zeta)}{{u_\star}^6T^3a^{3/2}}\left[F^3(1+H/\beta)^3\right]  \label{ALMOSTTHERE}		, \\
{u_\star}^6 & = & \frac{\kappa^2z^2\epsilon^2}{(1+d(-\zeta)^{2/3})^3} \label{USTARSQUARED}						,
\end{eqnarray}
where $F(T,p,q,\lambda_1,\lambda_2,C_{n_1}^2,C_{n_2}^2)$ and $H(T,p,q)$ are defined as

\begin{eqnarray}
F(T,p,q,\lambda_1,\lambda_2,C_{n_1}^2,C_{n_2}^2)& = &\frac{\mbox{sign}_1\sqrt{{C_{n_1}^2}}B_2-\mbox{sign}_2\sqrt{{C_{n_2}^2}}B_1}{A_1B_2-A_2B_1}	, \\
H(T,p,q)& = &E\left(\frac{0.61T}{\rho+0.61q}\right) 												.
\end{eqnarray}
We then combine Eqs. \ref{ALMOSTTHERE} and \ref{USTARSQUARED} to obtain a final equation in $\zeta$:

\begin{equation}
\zeta^3 = M(1-b\zeta)(1+d(-\zeta)^{2/3})^3 \label{cubicnonnormal}		,
\end{equation}
where 

\begin{equation}
M\equiv\frac{g^3z^2\kappa [F^3(1+H/\beta)^3]}{T^3\epsilon^2a^{3/2}}		,
\end{equation}
is determined directly from the source measurements. Here we note that the left-hand side is negative, and so the term in square brackets in $M$ is negative as well. 
From any set of measurements we know the sign of $A_1B_2-A_2B_1$, and we also know the values of the two terms that multiply the unknown signs.
Occasionally these relations are enough to determine all the signs; otherwise the signs remain ambiguous and they are evaluated from observations of the temperature and humidity stratification
as seen in \inlinecite{ANDREAS1989}.\newline

\begin{figure}
  \centering
  \includegraphics[width=11cm]{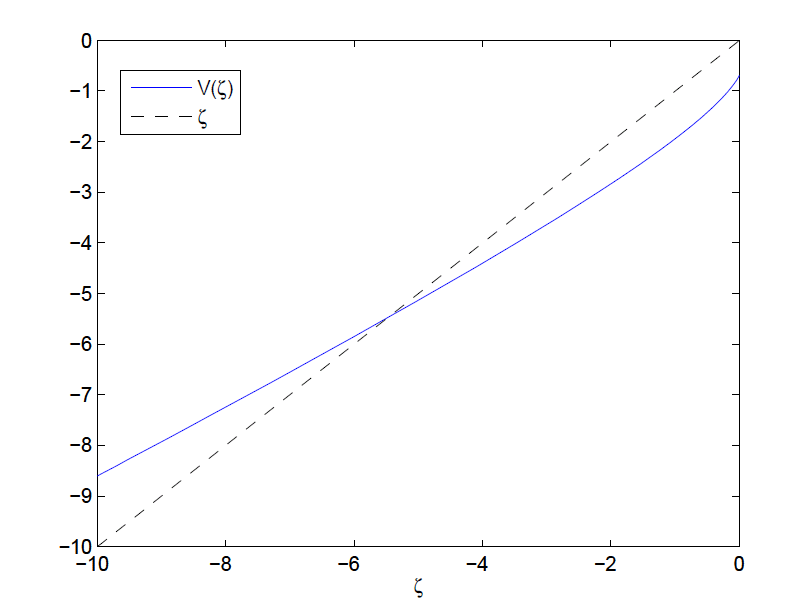}
  \caption[Visualization of the solution of Eq. \ref{cubicnonnormal} using fixed-point recursion.]
    {Visualization of the solution of Eq. \ref{cubicnonnormal} using fixed-point recursion, with $M=-1/3$. The function $\zeta=V(\zeta)$ is used, where \newline $V(\zeta)\equiv M^{1/3}(1-b\zeta)^{1/3}(1+d(-\zeta)^{2/3})$.
    Real roots of $M^{1/3}$ are chosen. The recursive series $[V(\zeta_{guess}),V(V(\zeta_{guess})),V(V(V(\zeta_{guess}))),V(V(V(V(\zeta_{guess}))))...]$ converges for any $\zeta_{guess}<0$.}
  \label{MVSZETAFIXED}
\end{figure}

\begin{figure}
  \centering
  \includegraphics[width=11cm]{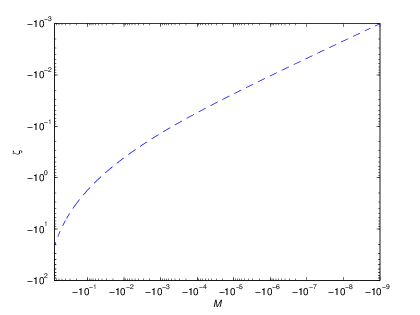}
  \caption[Solution of Eq. \ref{cubicnonnormal} using fixed-point recursion.]
    {Solution of Eq. \ref{cubicnonnormal} using fixed-point recursion on the function $\zeta=V(\zeta)$ where $V(\zeta)\equiv M^{1/3}(1-b\zeta)^{1/3}(1+d(-\zeta)^{2/3})$.
    Real roots of $M^{1/3}$ are chosen. Note that for $M=-1/3$, we have $\zeta\approx -5.5$ as in Fig. \ref{MVSZETAFIXED}. 
    Computational error was verified to be completely negligible with minimal running time involved.}
  \label{MVSZETA}
\end{figure}

Eq. \ref{cubicnonnormal} can be solved with a fixed-point recursive technique as illustrated in Fig. \ref{MVSZETAFIXED}.
The recursive function 

\begin{equation}
\zeta=V(\zeta)\equiv M^{1/3}(1-b\zeta)^{1/3}(1+d(-\zeta)^{2/3})
\end{equation}
is used.
A solution 
of Eq. \ref{cubicnonnormal} using fixed-point recursion is seen in Fig. \ref{MVSZETA}.\newline

A good estimate of the uncertainty in the derived variables that results from small
errors in source measurements is given by

\begin{equation}
 {\sigma_{f}} = \sum_{i=1}^{N}\left({\frac{\partial f}{\partial x_i}}\right)\sigma_{x_{s_i}} 
 + \sqrt{\sum_{i=1}^{N}\left({\frac{\partial f}{\partial x_i}}\right)^2{\sigma^2}_{x_{r_i}}} + \sigma_{f_{c}}	\label{errorprop}		,
\end{equation}
where the derived variable $f$ is a function of source measurement variables $x_{1},x_{2},...,x_{N}$ with respective systematic error $\sigma_{x_{s_1}},\sigma_{x_{s_2}},...,\sigma_{x_{s_N}}$ and 
with respective independent Gaussian distributed uncertainties with standard deviations $\sigma_{x_{r_1}},\sigma_{x_{r_2}},...,\sigma_{x_{r_N}}$ as seen in \inlinecite{TAYLOR}.
The numerical indices indicate different independent variables, such as $T$, $p$, or $z$, for example.
Computational error $f$ due to the inaccurate solution of the theoretical equations is represented by $\sigma_{f_{c}}$. The first and last terms in Eq. \ref{errorprop} represent an offset 
from the true solution (inaccuracy), whereas the central square-root term represents the breadth of uncertainty due to random error (imprecision). \newline

It is practical for the purpose of a sensitivity study 
to rewrite Eq. \ref{errorprop} as

\begin{equation}
 \frac{\sigma_{f}}{f} = \sum_{i=1}^{N}S_{f,x}\frac{\sigma_{x_{s_i}}}{x_{s_i}} 
 + \sqrt{\sum_{i=1}^{N}S_{f,x}^2\frac{{\sigma^2}_{x_{r_i}}}{{x_{r_i}}^2}} 	+ \frac{\sigma_{f_{c}}}{f}		,		\label{errorprop2}	
\end{equation}
where $S_{f,x}$ are unitless sensitivity functions defined by

\begin{equation}
S_{f,x} \equiv \frac{x}{f}\left(\frac{\partial f}{\partial x}\right)	.  \label{SENSITIVITYEQ}
\end{equation}
The sensitivity functions are each a measure of the portion of the error 
in the derived variable $f$ resulting from error on each individual source measurement $x$. In addition to the error on source measurement variables, we
can also recognize
that $a$, $b$ and $d$ have been resolved to some level of certainty by fitting field data. 
We thus treat them here in the same way as source measurements. \newline

In the application of Eqs. \ref{errorprop} and \ref{errorprop2}, we recognize the addition of the computational error $\sigma_{f_{c}}$.  
In previous field and sensitivity studies \cite{LAGOUARDE,DEBRUIN2002,SOLIGNAC,ANDREAS2012}, 
the full set of equations has been incorporated into 
a cyclically iterative algorithm which cycles through the full set of equations, allowing multiple variables to change. 
This numerical algorithm sometimes fails to converge, as demonstrated in \inlinecite{ANDREAS2012}.\newline%chi^2 increase by 1
%maybe include as appendix that cool code

The problem of resolving the uncertainty on the derived variables is a matter of identifying the magnitude and character of the source measurement uncertainties,
and then solving for the 
partial derivative terms in Eqs. \ref{errorprop} and \ref{SENSITIVITYEQ}. These derivatives are global\footnote{
Global partial derivatives are those which propagate from the dependent (derived) variable down to the independent (source measurement) variable through 
the entire tree
diagram, whereas local partial derivatives propagate as if the equation being differentiated were independent of the rest of the equations in the set. An alternative to direct evaluation
of global partial derivatives via the chain rule is
a total-differential expansion (where all derivatives are local) of each equation in the set. 
This approach can be used to solve for global partial derivatives by re-grouping all total-differential terms into one equation. Readers may refer to \inlinecite{CHAINRULE}.}
; that is, they take into account 
all the relationships in all of the relevant equations through which the variable $f$ is derived. Without an analytic solution of the set of coupled equations
we could either solve for the partial derivatives
through a total-differential expansion of each equation individually, followed by a re-grouping of all differential terms as seen in 
Andreas (1989; 1992)
%\inlinecite{ANDREAS1989,ANDREAS1992}, 
or we could use numerical 
error propagation techniques as in the 
Monte Carlo analysis of \inlinecite{MORONI} or as in the analysis of \inlinecite{SOLIGNAC}.\newline 

\begin{figure}
  \centering
  \includegraphics[width=11cm]{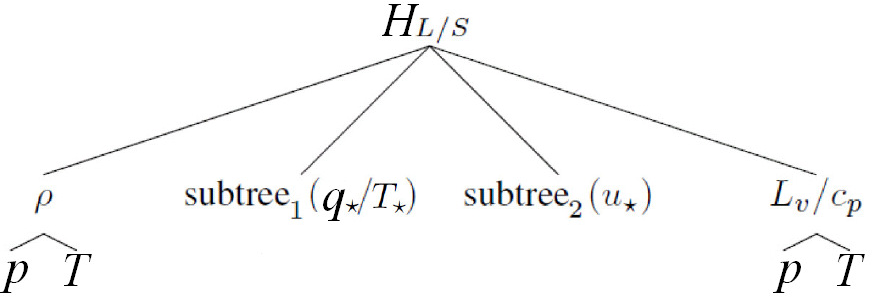}
  \caption[Variable inter-dependency tree diagram.]
    {Variable inter-dependency tree diagram for
    a two-wavelength measurement strategy inferring $H_{L/S}$ through path-averaged $u_\star$ and $q_\star/T_\star$ 
    measurements via scintillometer measurements 
    of $l_o$ and $C_n^2$ under unstable meteorological conditions ($\zeta<0$). Variables at the bottom of the tree are source measurements; all others are considered to be derived variables.
    The ``$/$'' symbol is meant to delineate between two independent tree diagrams. 
    Note that $H_L$ is not a direct function of $\rho$; this branch is for the convenience of including 
    $H_S$ since the rest of their tree diagrams are identical. Figs. \ref{SUBTREE1} and \ref{SUBTREE2} feature $\mbox{subtree}_1$ and $\mbox{subtree}_2$, respectively.}
  \label{TREE1}
\end{figure}

\begin{figure}
  \centering
  \includegraphics[width=14cm]{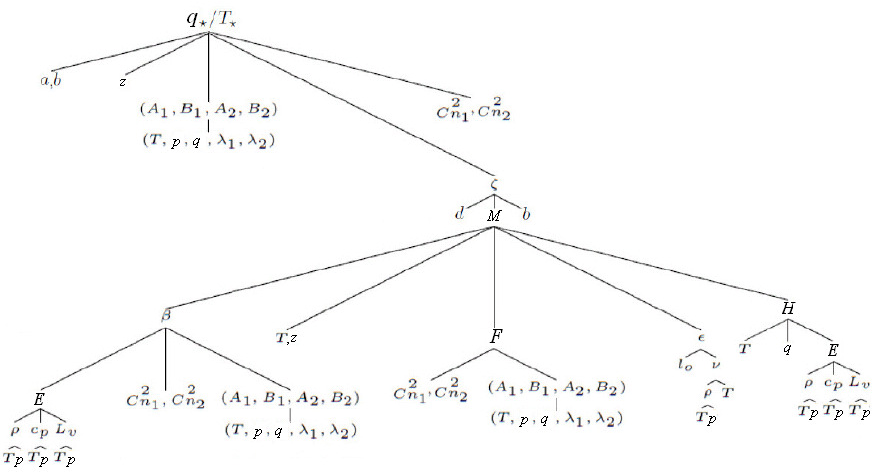}
  \caption[$\mbox{Subtree}_1$ of variable inter-dependency for $\zeta<0$.]
    {$\mbox{Subtree}_1$ of variable inter-dependency for $\zeta<0$. The main tree diagram is seen in Fig. \ref{TREE1}.}
  \label{SUBTREE1}
\end{figure}

\begin{figure}
  \centering
  \includegraphics[width=14cm]{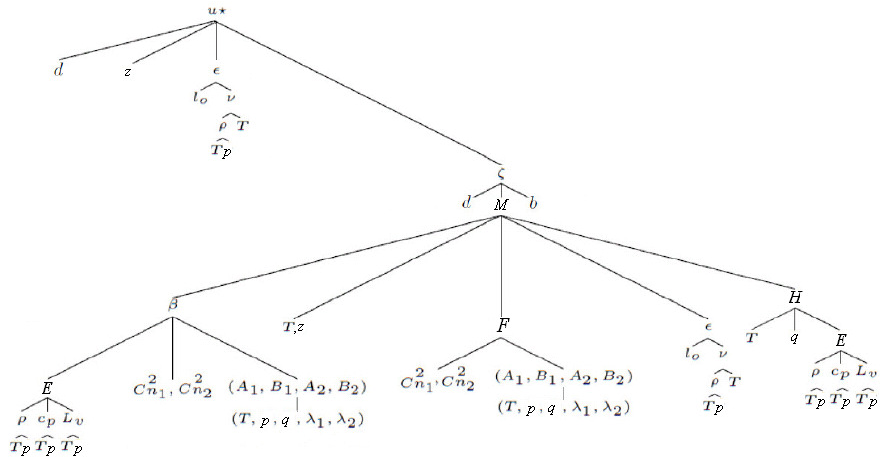}
  \caption[$\mbox{Subtree}_2$ of variable inter-dependency for $\zeta<0$.]
    {$\mbox{Subtree}_2$ of variable inter-dependency for $\zeta<0$. The main tree diagram is seen in Fig. \ref{TREE1}.}
  \label{SUBTREE2}
\end{figure}

We investigate 
inter-variable sensitivity analytically via Eq. \ref{SENSITIVITYEQ}, using Eq. \ref{cubicnonnormal} as a starting point. We use Eq. \ref{cubicnonnormal}
to determine the details of the variable inter-dependency to define 
our use of the chain rule.
A tree diagram representing the variable inter-dependency is broken into three parts shown in Figs. \ref{TREE1}, \ref{SUBTREE1}, and \ref{SUBTREE2}.\newline

%\EPSFIG[scale=0.3]{tree1R1.pdf}{Variable inter-dependency hierarchy for
%a two-wavelength measurement strategy inferring $H_{L/S}$ through path-averaged $u_\star$ and $q_\star/T_\star$ 
%measurements via scintillometer measurements 
%of $l_o$ and $C_n^2$ under unstable meteorological conditions ($\zeta<0$). Variables at the bottom of the tree are source measurements; all others are considered to be derived variables.
%The ``$/$'' symbol is meant to delineate between two independent tree diagrams. 
%Note that $H_L$ is not a direct function of $\rho$, the branch is for the convenience of including 
%$H_S$ since the rest of their tree diagrams are identical. Figures \ref{SUBTREE1} and \ref{SUBTREE2} feature $\mbox{subtree}_1$ and $\mbox{subtree}_2$, respectively.}{TREE1}
%

%\EPSFIG[scale=0.4]{subtree1R1.pdf}{$\mbox{Subtree}_1$ of variable inter-dependency hierarchy for $\zeta<0$. The main tree diagram is seen in Figure \ref{TREE1}.}{SUBTREE1}
%

%\EPSFIG[scale=0.4]{subtree2R1.pdf}{$\mbox{Subtree}_2$ of variable inter-dependency hierarchy for $\zeta<0$. The main tree diagram is seen in Figure \ref{TREE1}.}{SUBTREE2}
%

Eq. \ref{cubicnonnormal} can be reduced to a choice of two algebraic equations

\begin{eqnarray}
&&\alpha> 0, -\alpha^9  =  M(1+d\alpha^2)^3(1+ b\alpha^3) , \zeta=-\alpha^3, \frac{\partial \zeta}{\partial \alpha} = -3\alpha^2 < 0 	, 	\nonumber \\
\label{poly1} \\ 
&&\alpha< 0, \alpha^9  =  M(1+d\alpha^2)^3(1-b\alpha^3) , \zeta=\alpha^3, \frac{\partial \zeta}{\partial \alpha} = 3\alpha^2 > 0  	, \label{poly2}
\end{eqnarray}
with the substitution

\begin{equation}
\alpha^2\equiv(-\zeta)^{2/3}>0	.
\end{equation}
Galois theory implies that, since Eqs. \ref{poly1} and \ref{poly2} are ninth order, there is no way to write $\zeta=f(p,T,q,C_{n_1}^2,C_{n_2}^2,\lambda_1,\lambda_2,z,l_o)$
for any general values of $b$ and $d$, where $f$ is an explicit function of the source measurements \cite{GALOIS}.
It is thus simplest to extract 
$\left(\frac{\partial \zeta}{\partial M}\right)$ by implicit differentiation of Eq. \ref{cubicnonnormal}; the results are in
given in Appendix \ref{APPNDX2a}.

\section{Results: Derivation of Sensitivity Functions}\label{RESULTS}

Following the solution method described above, we solve for global
partial derivative terms in Eqs. \ref{errorprop} and \ref{SENSITIVITYEQ}
through use of the general chain rule guided by the variable inter-dependency 
tree diagrams seen in Figs. \ref{TREE1}, \ref{SUBTREE1} and \ref{SUBTREE2}. We will obtain sensitivity functions of the sensible heat flux $H_S$ and the momentum flux $\tau$ as functions of 
$z$ and $\epsilon$. From Eqs. \ref{SENSIBLEHEAT}, \ref{LOEPSILON} and \ref{SENSITIVITYEQ} we have

\begin{eqnarray}
S_{H_S,\epsilon}&=&S_{T_\star,\epsilon}+S_{u_\star,\epsilon}=-\frac{1}{4}S_{H_S,l_o}	\label{sensiblesensitivityepsilon}		,\\
S_{H_S,z}&=&S_{T_\star,z}+S_{u_\star,z}		 		\label{sensiblesensitivityz}		,
\end{eqnarray}
and from Eqs. \ref{MOMENTUM}, \ref{LOEPSILON} and \ref{SENSITIVITYEQ}, we have

\begin{eqnarray}
S_{\tau,\epsilon}&=&2 S_{u_\star,\epsilon}=-\frac{1}{4}S_{\tau,l_o}	\label{momentumsensitivityepsilon}		,\\
S_{\tau,z}&=&2 S_{u_\star,z}		 		\label{momentumsensitivityz} 		,
\end{eqnarray}
thus we seek solutions for $S_{T_\star,z}$, $S_{u_\star,z}$, $S_{T_\star,\epsilon}$, and $S_{u_\star,\epsilon}$.\newline

\begin{figure}
  \centering
  \includegraphics[width=11cm]{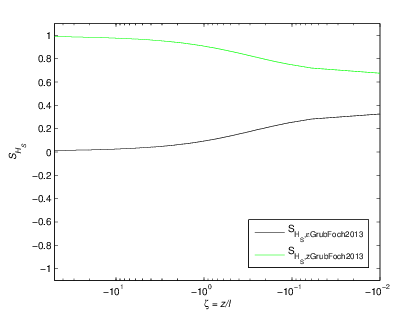}
  \caption[Sensitivity functions for $H_S$ with regards to measurements of $\epsilon$ and $z$.]
    {Sensitivity functions for $H_S$ with regards to measurements of $\epsilon$ and $z$ in the path-averaged $u_\star$ scintillation measurement, for unstable conditions corresponding to $\zeta<0$.}
  \label{SHSFIG}
\end{figure}

We first obtain $S_{T_\star,\epsilon}$ with guidance from the tree diagram depicted in Fig. \ref{SUBTREE1}:

\begin{equation}
S_{T_\star,\epsilon} = \frac{\epsilon}{T_\star} \left(\frac{\partial T_\star}{\partial \zeta}\right)\left(\frac{\partial \zeta}{\partial M}\right)\left(\frac{\partial M}{\partial \epsilon}\right)	. \label{STSTAREPSILONEQ}
\end{equation}
The individual terms of Eq. \ref{STSTAREPSILONEQ} are given in Appendices \ref{APPNDX2a} and \ref{APPNDX2b}. Combining them, we obtain

\begin{equation}
S_{T_\star,\epsilon} = \frac{1}{3}\left(\frac{2b\zeta(-\zeta)^{1/3}(1+d(-\zeta)^{2/3})}{(3-2b\zeta)(1+d(-\zeta)^{2/3})(-\zeta)^{1/3}+2d\zeta(1-b\zeta)}\right)	. \label{Ststarepsilon}
\end{equation}
We now obtain $S_{T_\star,z}$:

\begin{equation}
S_{T_\star,z} = \frac{z}{T_\star} \left[\left(\frac{\partial T_\star}{\partial z}\right)_\zeta + \left(\frac{\partial T_\star}{\partial \zeta}\right)_z \left(\frac{\partial \zeta}{\partial M}\right)\left(\frac{\partial M}{\partial z}\right)\right] . \label{STSTARZEQ}
\end{equation}
The individual terms of Eq. \ref{STSTARZEQ} are developed in Appendices \ref{APPNDX2a} and \ref{APPNDX2c}. Combining them, we obtain

\begin{equation}
S_{T_\star,z} = \frac{1}{3}\left[1-\left(\frac{2b\zeta(-\zeta)^{1/3}(1+d(-\zeta)^{2/3})}{(3-2b\zeta)(1+d(-\zeta)^{2/3})(-\zeta)^{1/3}+2d\zeta(1-b\zeta)}\right)\right] . \label{Ststarz}
\end{equation}
We now obtain $S_{u_\star,\epsilon}$ with guidance from the tree diagram depicted in Fig. \ref{SUBTREE2}. We have

\begin{equation}
S_{u_\star,\epsilon} = \frac{\epsilon}{u_\star}\left[\left(\frac{\partial u_\star}{\partial \epsilon}\right)_\zeta
+ \left(\frac{\partial u_\star}{\partial \zeta}\right)_\epsilon \left(\frac{\partial \zeta}{\partial M}\right)
\left(\frac{\partial M}{\partial \epsilon}\right)   \right]  . \label{SUSTAREPSILONEQ}
\end{equation}
The individual terms in Eq. \ref{SUSTAREPSILONEQ} are developed in Appendices \ref{APPNDX2a} and \ref{APPNDX2d}. Combining them, we obtain

\begin{equation}
S_{u_\star,\epsilon} = \frac{1}{3}\left[ 1 - \left(\frac{2d\zeta(1-b\zeta)}{(3-2b\zeta)(1+d(-\zeta)^{2/3})(-\zeta)^{1/3}+2d\zeta(1-b\zeta)}\right)\right] . \label{Sustarepsilon}
\end{equation}
We now obtain $S_{u_\star,z}$. We have

\begin{equation}
S_{u_\star,z} = \frac{z}{u_\star}\left[\left(\frac{\partial u_\star}{\partial z}\right)_\zeta
+ \left(\frac{\partial u_\star}{\partial \zeta}\right)_z \left(\frac{\partial \zeta}{\partial M}\right)
\left(\frac{\partial M}{\partial z}\right)   \right] . \label{SUSTARZEQ}
\end{equation}
The individual terms in Eq. \ref{SUSTARZEQ} are developed in Appendices \ref{APPNDX2a} and \ref{APPNDX2e}. Combining them we obtain

\begin{equation}
%S_{u_\star,z} & = & \frac{(1+d(-\zeta)^{2/3})+6d(-\zeta)^{1/3}M\frac{\partial \alpha}{\partial M}}{3(1+d(-\zeta)^{2/3})} \label{SSSz2} \\
S_{u_\star,z} = \frac{1}{3}\left[ 1 + \left(\frac{2d\zeta(1-b\zeta)}{(3-2b\zeta)(1+d(-\zeta)^{2/3})(-\zeta)^{1/3}+2d\zeta(1-b\zeta)}\right)\right] . \label{Sustarz}
\end{equation}
Combining our results in Eqs. \ref{STSTAREPSILONEQ}, \ref{STSTARZEQ},  \ref{SUSTAREPSILONEQ}, and \ref{SUSTARZEQ}, we can obtain $S_{H_S,\epsilon}$ and $S_{H_S,z}$ from
Eqs. \ref{sensiblesensitivityepsilon} and \ref{sensiblesensitivityz}; the results are seen in Fig. \ref{SHSFIG}.\newline

\begin{figure}
  \centering
  \includegraphics[width=11cm]{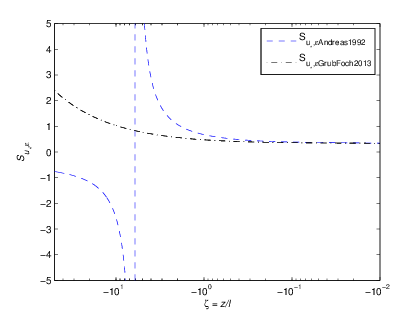}
  \caption[Sensitivity function for $u_\star$ with regards to measurements of $\epsilon$.]
    {Sensitivity function for $u_\star$ with regards to measurements of $\epsilon$ in the path-averaged $u_\star$ scintillation measurement. Results from \inlinecite{ANDREAS1992} 
    are plotted (denoted there as $S_\epsilon$) along with Eq. \ref{Sustarepsilon} derived here for $\zeta<0$.}
  \label{SUSTAREPSILONBROADS}
\end{figure}

\begin{figure}
  \centering
  \includegraphics[width=11cm]{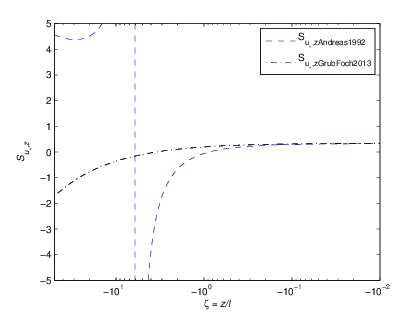}
    \caption[Sensitivity function for $u_\star$ with regards to measurements of $z$.]
    {Sensitivity function for $u_\star$ with regards to measurements of $z$ in the path-averaged $u_\star$ scintillation measurement. Results from \inlinecite{ANDREAS1992} 
    are plotted (denoted there as $S_{zz}$) along with Eq. \ref{Sustarz} derived here for $\zeta<0$.}
  \label{SUSTARZBROADS}
\end{figure}

The absolute value of our results for $S_{H_S,l_o}$ given by Eqs. \ref{sensiblesensitivityepsilon}, \ref{Ststarepsilon} and \ref{Sustarepsilon}
is similar to the sensitivity multiplier found in \inlinecite{MORONI} as seen in their Fig. 10.
The absolute value of our result of $S_{\tau,l_o}$ given by Eqs. \ref{momentumsensitivityepsilon} and \ref{Sustarepsilon} is also compatible with the results of \inlinecite{MORONI} seen in their Fig. 9.
However, our result for $S_{u_\star,\epsilon}$ in Eq. \ref{Sustarepsilon} differs from that obtained in \inlinecite{ANDREAS1992}
as seen in Fig. \ref{SUSTAREPSILONBROADS}.
%\EPSFIG[scale=0.9]{Sustarepsilonbroad.pdf}{Sensitivity function for $u_\star$ with regards to measurements of $\epsilon$ in the path-averaged $u_\star$ scintillation measurement. Results from \inlinecite{ANDREAS1992} 
%are plotted (there denoted $S_\epsilon$) along with Equation (\ref{Sustarepsilon}) derived here for $\zeta<0$.}{SUSTAREPSILONBROADS}
%
Similarly, our result for $S_{u_\star,z}$ in Eq. \ref{Sustarz} differs from that obtained in \inlinecite{ANDREAS1992} as seen in Fig. \ref{SUSTARZBROADS}.

%\EPSFIG[scale=0.9]{Sustarzbroad.pdf}{Sensitivity function for $u_\star$ with regards to measurements of $z$ in the path-averaged $u_\star$ scintillation measurement. Results from \inlinecite{ANDREAS1992} 
%are plotted (there denoted $S_{zz}$) along with results derived here in Equation (\ref{Sustarz}) for $\zeta<0$.}{SUSTARZBROADS}
%

\section{Discussion}\label{DISCUSSION}

The reason for the difference between our results and those of \inlinecite{ANDREAS1992} 
in Figs. \ref{SUSTAREPSILONBROADS} and \ref{SUSTARZBROADS}
can be seen to have arisen in Eqs. A.7 and A.10 of \inlinecite{ANDREAS1992} . Even though there is a typographical error in Eq. A.7 in the application of the product rule 
(it should be

\begin{equation}
\frac{\partial \epsilon}{\partial u_\star} = \frac{3{u_\star}^2}{\kappa z}\phi_\epsilon(\zeta)+\frac{{u_\star}^3}{z\kappa}\frac{\partial \phi_\epsilon}{\partial \zeta}\frac{\partial \zeta}{\partial u_\star} , \label{TYPO}
\end{equation}
where the second term contained $u_\star^2$ originally),
this is not the origin of the reason since the result in Eq. A.8 follows from the modified Eq. A.7. The reason
is found to be that Eqs. A.7 and A.8 are not differentiated locally with respect to Eq. 1.3 of \inlinecite{ANDREAS1992} 
as they should be in a total-differential expansion.
The local derivative is

\begin{equation}
\frac{\partial \epsilon}{\partial u_\star} = \frac{\partial}{\partial u_\star}\left(\frac{u_\star^3}{\kappa z}\phi_\epsilon(\zeta)\right) = \frac{3{u_\star}^2}{\kappa z}\phi_\epsilon(\zeta) = \frac{3\epsilon}{u_\star} , \label{LOCALUSTAR}
\end{equation}
keeping $\zeta$ constant regardless of the relationship between $\zeta$ and $u_\star$. The relationship between $\zeta$ and $u_\star$
is taken into account when we re-group 
the full set of locally expanded equations (which are coupled in $\zeta$ and $u_\star$). 
The second term on the right-hand side of Eq. \ref{TYPO} and Eq. A.7 of
\inlinecite{ANDREAS1992} is thus not necessary and does not appear in Eq. \ref{LOCALUSTAR}. 
Taking into account the relationship between $\zeta$ and $u_\star$ via the chain rule
is appropriate for direct evaluation of global derivatives, but not in individual derivatives of
a total-differential expansion of the full set of equations.
Eqs. A.10 and A.11 of \inlinecite{ANDREAS1992} have the same issues of not being differentiated locally with respect to Eq. 1.3 
of \inlinecite{ANDREAS1992}. The local derivative there is 

\begin{equation}
\frac{\partial \epsilon}{\partial z} = -\frac{\epsilon}{z} . \label{localz}
\end{equation}
A re-analysis of the \inlinecite{ANDREAS1992} differential expansion including the local derivatives in Eqs. \ref{LOCALUSTAR}
and \ref{localz} is reproduced in Appendix \ref{EXPANSION}; the results for $S_{u_\star,\epsilon}$ and 
$S_{u_\star,z}$ are identical to those found here in Eqs. \ref{SUSTAREPSILONEQ} and \ref{SUSTARZEQ}.
Note that the left-hand side of Eq. \ref{FIFY2} contains the terms $(S_{u_\star}-2)$ and $(S_{z}+1)$ instead of
$(S_{u_\star}-4)$ and $(S_{z}+2)$ as in Eq. A.16 of 
\inlinecite{ANDREAS1992}. These differences also influence the \inlinecite{ANDREAS1992} sensitivity functions for ${C_n^2}_1$ and ${C_n^2}_2$.\newline

The technique presented here for the direct evaluation of partial derivatives
can be applied to evaluate sensitivity functions for other variables involved in this scintillometer strategy
for both stable and unstable conditions, however we will now focus on the implications of our results on other previous studies.
Another instance where we found divergence in results is in the study of \inlinecite{HARTOGENSIS2003} 
where $S_{H_S,z}$ in Eq. A2 and Fig. A1 should be the same as the results of \inlinecite{ANDREAS1989} in Fig. 4, regardless 
of the differences between a single and double wavelength strategy. Note that in \inlinecite{ANDREAS1989}, for $\zeta = 0$, it was found that

\begin{equation}
S_{H_S,z}(0)=S_{T_\star,z}(0)= 1/3	,
\end{equation}
for a 
scintillometer strategy involving independent $u_\star$ measurements, whereas a value of $1/2$ was found
in \inlinecite{HARTOGENSIS2003}.
The issue here is not due to the differences in scintillation strategies (note that the Businger-Dyer relation is ignored in the sensitivity study of \inlinecite{HARTOGENSIS2003}). 
The issue is that Eq. A1 of \inlinecite{HARTOGENSIS2003} is coupled to Eqs. 5-6 of \inlinecite{HARTOGENSIS2003} in $L$.
In the derivation of Eq. A1, \inlinecite{HARTOGENSIS2003} essentially have considered $Z_{LAS}$ to be the same $z$ as in \inlinecite{ANDREAS1989}, and they have considered similar equations that 
assume an independent $u_\star$ measurement (Eq. 7 of \inlinecite{HARTOGENSIS2003} is ignored). Including the coupling of Eq. 7 of \inlinecite{HARTOGENSIS2003} (the Businger-Dyer relation)
in $L$ adds complication; 
however if we continue to assume an independent $u_\star$ measurement, we achieve the same results 
as in \inlinecite{ANDREAS1989}, viz:

\begin{equation}
S_{H_S,z}=S_{T_\star,z}=\frac{1-2b\zeta}{3-2b\zeta}\neq\frac{1-2b\zeta}{2-2b\zeta}=\frac{z}{H_S}\left(\frac{\partial H_S}{\partial z}\right)_L 		.
\end{equation}
A similar example is in the analysis of \inlinecite{HARTOGENSIS}, when the sensitivity of
$u_\star$ to $l_o$ is being examined. Eq. 13 of \inlinecite{HARTOGENSIS} is not a ``direct'' relation of $u_\star$ to source measurements, since $L$ is a derived variable. There is coupling to $L$ and thus we may investigate the sensitivity
with

\begin{equation}
\left(\frac{\partial u_\star}{\partial l_o}\right) = \left(\left(\frac{\partial u_\star}{\partial \epsilon}\right)_\zeta+
\left(\frac{\partial u_\star}{\partial \zeta}\right)\left(\frac{\partial \zeta}{\partial M}\right)\left(\frac{\partial M}{\partial \epsilon}\right)
\right)\left(\frac{\partial \epsilon}{\partial l_0}\right) 	,
\end{equation}
where $M$ is modified for the single scintillometer $l_o$ and $C_n^2$ strategy. Also in \inlinecite{HARTOGENSIS}, it is stated
that errors in ${C_T^2}$ are attenuated in deriving $\theta_\star$ (here denoted $T_\star$) due to the square-root dependence; however 
we can go a step further by realizing that Eq. 9 of \inlinecite{HARTOGENSIS} is not yet decoupled from $L$. 
As follows from our analysis applied to the case considered in \inlinecite{HARTOGENSIS} (modifying Fig. \ref{SUBTREE1} for a single-wavelength strategy), we obtain

\begin{equation}
\left(\frac{\partial T_\star}{\partial {C_T^2}}\right) =
\left(\frac{\partial T_\star}{\partial {C_T^2}}\right)_\zeta+\left(\frac{\partial T_\star}{\partial \zeta}\right)
\left(\frac{\partial \zeta}{\partial M}\right)\left(\frac{\partial M}{\partial {C_T^2}}\right) 	.
\end{equation}
Note that there may be no way to actually obtain ``direct'' relationships between the source 
measurements and the derived variables if the implicit equation in $\zeta$ (such as Eq. \ref{cubicnonnormal}) is fifth order or higher. \newline

\section{Conclusions}\label{CONCLUSIONS}

A new method of deriving sensitivity functions for $l_o$ and $C_n^2$ scintillometer measurements of turbulent fluxes has been produced by mapping 
out the variable inter-dependency and solving for partial derivatives with the chain rule. We have bypassed the need for an explicit solution 
to the theoretical equations by including one implicit differentiation step on Eq. \ref{cubicnonnormal}, which is 
a bottleneck on the tree diagrams seen in Figs. \ref{SUBTREE1} and \ref{SUBTREE2}.
This allows for the evaluation of
sensitivity functions that are useful not only for optimizing the measurement strategy and selecting the most ideal wavelengths,
but the closed, compact form of sensitivity functions produced using the method presented here is convenient
to incorporate into computer code for the analysis of data. 
It is noteworthy that the actual functional relations change at $z/L=0$, which corresponds to neutral conditions.
Thus, for any set of source measurements we should calculate the set of all derived variables and their respective uncertainties assuming both stable and unstable conditions.  
If errors 
on $z/L$ overlap with $z/L=0$ for either stability regime, we should then consider the combined range of errors. \newline

In addition to the source measurements, the empirical parameters $a$, $b$ and $d$ have been included in the tree diagrams. 
Future study should quantify the sensitivity of derived variables to these parameters.
In considering errors on the empirical parameters or on other source measurements such as $T$, 
a total-differential expansion such as in 
Andreas (1989; 1992)
%\inlinecite{ANDREAS1989,ANDREAS1992} 
may become intractable, whereas an analysis of the type presented here remains compact.\newline

Results obtained here have resolved some issues in the previous literature.
For example, we have confirmed the conclusion of \inlinecite{MORONI} that $l_o$ and $C_n^2$ scintillometers can obtain fairly precise measurements of turbulent fluxes. 
In the range of $-1\leq\zeta\leq-0.01$, the results derived here for $S_{u_\star,\epsilon}$ and $S_{u_\star,z}$ 
are similar to those in \inlinecite{ANDREAS1992}; 
however for $\zeta < -1$ the separate results differ greatly in both magnitude 
and in the shape of the curves as seen in Figs. \ref{SUSTAREPSILONBROADS} and \ref{SUSTARZBROADS}. 
These sensitivity functions in \inlinecite{ANDREAS1992} contain singularities near $\zeta \approx -6$; this effectively implies that it is impossible to resolve 
$u_\star$ in this stability regime. The sensitivity functions derived here demonstrate a small magnitude for typical values of $\zeta$ including the range $-10<\zeta<-1$.
The sensitivities of the sensible heat flux to uncertainties in $\epsilon$ and $z$ are found in Eqs. \ref{sensiblesensitivityepsilon} and 
\ref{sensiblesensitivityz} and are seen in Fig. \ref{SHSFIG}; they are compatible with the results of \inlinecite{MORONI} and they imply that, with optimal wavelengths, we can arrive at 
reasonably precise measurements of path-averaged turbulent fluxes and friction velocity.\newline

An advantageous byproduct of having reduced the system of equations into a single equation in a single unknown is that the error in the actual computation of the derived variables can be 
essentially eliminated, or it can be estimated. Eqs. \ref{poly1} and \ref{poly2} are polynomials; numerical methods for their accurate solution are well established. 
Using fixed-point recursion, the maximum computational error can be resolved, and monotonic convergence can be guaranteed as seen in \inlinecite{ITERATIVE} and more recently in \inlinecite{FIXEDPOINT}.\newline

In contrast, 
the classical iterative algorithm 
%\cite{ANDREAS1989,HARTOGENSIS2003,SOLIGNAC,ANDREAS2012}
(Andreas, 1989; 2012; Hartogensis, 2003; Solignac, 2009)
may diverge or alternate about a potential solution. At worst, techniques such as the classical algorithm
may stop at a ``bottleneck'' and converge to a false solution as illustrated in \inlinecite{PRESSNUM}. In their section on 
non-linear coupled equations, it is stated:\newline

\textit{``We make an extreme, but wholly defensible, statement: there are no good, general} (numerical) \textit{methods for solving systems of more than one non-linear equation. Furthermore, 
it is not hard to see why (very likely), 
there never will be any good, general} (numerical) \textit{methods...''}\newline

In \inlinecite{HILLSINGLEEQ}, similar one-dimensional iterative methods of numerical computation of $\zeta$ were used to eliminate computational error,
however the fixed-point algorithm we have presented converges for any $\zeta_{guess}$ (with the correct sign).
We argue that at least some of the spread of data in Figs. 5 and 6 in \inlinecite{ANDREAS2012} may be due to 
computational uncertainty as well as the incorporation of $T_\star$, $L$, and $u_\star$ measured at the 
scale of an eddy covariance system's footprint while being forced to assume that they are representative of the beam path scale. 
The scatter in these plots may not be entirely due to unreliable $l_o$ and $C_n^2$ measurements.\newline

Future expansions of the sensitivity analysis presented here may focus on taking into account field sites with 
heterogeneous terrain and variable topography. For stationary turbulence with beams above the blending height, 
the line integral formulation for effective beam height given by Eq. B2 in \inlinecite{HARTOGENSIS2003} 
and Eqs. 10-12 in \inlinecite{KLEISSL2008} could be incorporated. 
Two-dimensional footprint analyses involving surface integrals that take into account variable roughness length
and wind direction 
as in \inlinecite{MEIJNINGERPWF2002} and in \inlinecite{LIU} may be incorporated for flat terrain that is heterogeneous enough to force 
the scintillometer beam to be below the blending height
\cite{WIERINGA1986,MASON1987}. Further theoretical developments may be anticipated that take into account both heterogeneity
and variable topography. It is hoped that the general mathematical approach presented here
can help to keep track of uncertainty for 
any scintillometer application, as well as to eliminate the byproducts of iteration.

\acknowledgements
The authors thank the Geophysical Institute at the University of Alaska Fairbanks for its support, Derek Starkenburg and 
Peter Bieniek for assistance with editing, two anonymous reviewers, one in particular, for very helpful comments.
In addition, the authors thank 
Flora Grabowska of the Mather library for her determination in securing funding 
for open access fees. GJ Fochesatto was partially supported by the Alaska Space Grant NASA-EPSCoR program award number NNX10N02A.

\appendix
\section{Relations between $M$ and $\zeta$}\label{APPNDX2a}

\begin{eqnarray}
M & = & \frac{\zeta^3}{(1+d(-\zeta)^{2/3})^3(1-b\zeta)}  ,\\
\left(\frac{\partial \zeta}{\partial M}\right) & = & \left(\frac{(1-b\zeta)(1+d(-\zeta)^{2/3})^3}{3\zeta^2+M[2d(1-b\zeta)(1+d(-\zeta)^{2/3})^2(-\zeta)^{-1/3}+b(1+d(-\zeta)^{2/3})^3]}\right) ,\nonumber \\
\\
M\left(\frac{\partial \zeta}{\partial M}\right) & = & \left( \frac{\zeta(1-b\zeta)(1+d(-\zeta)^{2/3})}{(3-2b\zeta)(1+d(-\zeta)^{2/3})+2d\zeta(-\zeta)^{-1/3}(1-b\zeta)} \right).
\end{eqnarray}

\section{Individual terms in $S_{T_\star,\epsilon}$ for unstable conditions ($\zeta<0$)}\label{APPNDX2b}

\begin{eqnarray}
\left(\frac{\partial T_\star}{\partial \zeta}\right) & = & T_\star\left(\frac{-b}{3(1-b\zeta)}\right),  \\
\left(\frac{\partial M}{\partial \epsilon}\right) & = & -2M/\epsilon .
\end{eqnarray}

\section{Individual terms in $S_{T_\star,z}$ for unstable conditions ($\zeta<0$)}\label{APPNDX2c}

\begin{eqnarray}
\left(\frac{\partial T_\star}{\partial z}\right)_\zeta & = & \frac{T_\star}{3z} ,\\
\left(\frac{\partial T_\star}{\partial \zeta}\right)_z & = & T_\star\left(\frac{-b}{3(1-b\zeta)}\right),  \\
\left(\frac{\partial M}{\partial z}\right) & = & 2M/z .
\end{eqnarray}

\section{Individual terms in $S_{u_\star,\epsilon}$ for unstable conditions ($\zeta<0$)}\label{APPNDX2d}

\begin{eqnarray}
\left(\frac{\partial u_\star}{\partial \epsilon}\right)_\zeta & = & \frac{u_\star}{3\epsilon} , \\
\left(\frac{\partial u_\star}{\partial \zeta}\right)_\epsilon & = & u_\star \left(\frac{d}{3(1+d(-\zeta)^{2/3})(-\zeta)^{1/3}}\right) , \\ 
\left(\frac{\partial M}{\partial \epsilon}\right) & = & -2M/\epsilon .
\end{eqnarray}

\section{Individual terms in $S_{u_\star,z}$ for unstable conditions ($\zeta<0$)}\label{APPNDX2e}

\begin{eqnarray}
\left(\frac{\partial u_\star}{\partial z}\right)_\zeta & = & \frac{u_\star}{3z} , \\
\left(\frac{\partial u_\star}{\partial \zeta}\right)_z & = & u_\star \left(\frac{d}{3(1+d(-\zeta)^{2/3})(-\zeta)^{1/3}}\right) , \\ 
\left(\frac{\partial M}{\partial z}\right) & = & 2M/z .
\end{eqnarray}

\section{Total differential expansion as in Andreas (1992) for unstable conditions ($\zeta<0$)}\label{EXPANSION}

Here we reproduce the analysis of \inlinecite{ANDREAS1992}. Subscripts indicate the equation that is being differentiated locally.
The coupled equations are

\begin{eqnarray}
\zeta = \frac{zgk}{u_\star^2 T}(T_\star+\frac{0.61T}{\rho+0.61q}q_\star)											,\label{ze} \\
\epsilon = \frac{u_\star^3}{\kappa z}\phi(\zeta) = \frac{u_\star^3}{\kappa z}(1+d(-\zeta)^{2/3})^{3/2} 							,\label{ep} \\
T_\star=\frac{(1-b\zeta)^{1/3}z^{1/3}}{\sqrt(a)}\left(\frac{\mbox{sign}_1\sqrt{C_{n_1}^2}B_2-\mbox{sign}_2\sqrt{C_{n_2}^2}B_1}{A_1B_2-A_2B_1}\right)	,\label{te} \\
q_\star=\frac{(1-b\zeta)^{1/3}z^{1/3}}{\sqrt(a)}\left(\frac{\mbox{sign}_2\sqrt{C_{n_2}^2}A_1-\mbox{sign}_1\sqrt{C_{n_1}^2}A_2}{A_1B_2-A_2B_1}\right)	.\label{qe}
\end{eqnarray}
We expand Eqs. \ref{ze} and \ref{ep} as

\begin{eqnarray}
d\zeta=\left(\frac{\partial \zeta}{\partial z}\right)_{\ref{ze}}dz+\left(\frac{\partial \zeta}{\partial T_\star}\right)_{\ref{ze}}dT_\star+\left(\frac{\partial \zeta}{\partial q_\star}\right)_{\ref{ze}}dq_\star 		,\\
d\epsilon=\left(\frac{\partial \epsilon}{\partial u_\star}\right)_{\ref{ep}}du_\star+\left(\frac{\partial \epsilon}{\partial z}\right)_{\ref{ep}}dz+\left(\frac{\partial \epsilon}{\partial \zeta}\right)_{\ref{ep}}d\zeta 	.
\end{eqnarray}
Combining these, we obtain

\begin{eqnarray}
d\epsilon=\left[\left(\frac{\partial \epsilon}{\partial u_\star}\right)_{\ref{ep}}+\left(\frac{\partial \epsilon}{\partial \zeta}\right)_{\ref{ep}}\left(\frac{\partial \zeta}{\partial u_\star}\right)_{\ref{ze}}\right]du_\star \nonumber \\
+\left[\left(\frac{\partial \epsilon}{\partial z}\right)_{\ref{ep}}+\left(\frac{\partial \epsilon}{\partial \zeta}\right)_{\ref{ep}}\left(\frac{\partial \zeta}{\partial z}\right)_{\ref{ze}}\right]dz \nonumber \\
+\left(\frac{\partial \epsilon}{\partial \zeta}\right)_{\ref{ep}}\left(\frac{\partial \zeta}{\partial T_\star}\right)_{\ref{ze}}dT_\star \nonumber \\
+\left(\frac{\partial \epsilon}{\partial \zeta}\right)_{\ref{ep}}\left(\frac{\partial \zeta}{\partial q_\star}\right)_{\ref{ze}}dT_\star 	,\\
\nonumber \\
\nonumber \\
\frac{d\epsilon}{\epsilon}=\frac{u_\star}{\epsilon}\frac{du_\star}{u_\star}
\left[\left(\frac{\partial \epsilon}{\partial u_\star}\right)_{\ref{ep}}+\left(\frac{\partial \epsilon}{\partial \zeta}\right)_{\ref{ep}}\left(\frac{\partial \zeta}{\partial u_\star}\right)_{\ref{ze}}\right]  \nonumber \\
+\frac{z}{\epsilon}\frac{dz}{z}\left[\left(\frac{\partial \epsilon}{\partial z}\right)_{\ref{ep}}+\left(\frac{\partial \epsilon}{\partial \zeta}\right)_{\ref{ep}}\left(\frac{\partial \zeta}{\partial z}\right)_{\ref{ze}}\right] \nonumber \\
+\frac{T_\star}{\epsilon}\frac{dT_\star}{T_\star}\left(\frac{\partial \epsilon}{\partial \zeta}\right)_{\ref{ep}}\left(\frac{\partial \zeta}{\partial T_\star}\right)_{\ref{ze}} \nonumber \\ 
+\frac{q_\star}{\epsilon}\frac{dq_\star}{q_\star}\left(\frac{\partial \epsilon}{\partial \zeta}\right)_{\ref{ep}}\left(\frac{\partial \zeta}{\partial q_\star}\right)_{\ref{ze}}	,
\end{eqnarray}
where the local derivatives are given by

\begin{eqnarray}
\left(\frac{\partial \epsilon}{\partial u_\star}\right)_{\ref{ep}}= \frac{3\epsilon}{u_\star} ,\\
\left(\frac{\partial \zeta}{\partial u_\star}\right)_{\ref{ze}}= \frac{-2\zeta}{u_\star} ,\\
\left(\frac{\partial \epsilon}{\partial \zeta}\right)_{\ref{ep}}= \frac{\epsilon}{\phi(\zeta)}\frac{\partial \phi}{\partial \zeta}(\zeta) ,\\
\left(\frac{\partial \epsilon}{\partial z}\right)_{\ref{ep}}=-\frac{\epsilon}{z} ,\\
\left(\frac{\partial \zeta}{\partial z}\right)_{\ref{ze}}=\frac{\zeta}{z} ,\\
\zeta_T\equiv\frac{zg\kappa}{u_\star^2 T}T_\star ,\\
\zeta_q\equiv\frac{zg\kappa}{u_\star^2 T}\left(\frac{0.61T}{\rho+0.61q}\right)q_\star ,\\
\zeta=\zeta_T+\zeta_q ,\\
\left(\frac{\partial \zeta}{\partial T_\star}\right)_{\ref{ze}}=\frac{\zeta_T}{T_\star} ,\\
\left(\frac{\partial \zeta}{\partial q_\star}\right)_{\ref{ze}}=\frac{\zeta_q}{q_\star} .
\end{eqnarray}
Thus the expansion becomes

\begin{eqnarray}
\frac{d\epsilon}{\epsilon}=\frac{du_\star}{u_\star}\left(3-\frac{2\zeta}{\phi(\zeta)}\frac{\partial \phi}{\partial \zeta}(\zeta)\right) \nonumber \\
+\frac{dz}{z}\left(-1+\frac{\zeta}{\phi(\zeta)}\frac{\partial \phi}{\partial \zeta}(\zeta)\right)								\nonumber \\
+\frac{dT_\star}{T_\star}\frac{\zeta_T}{\phi(\zeta)}\frac{\partial \phi}{\partial \zeta}(\zeta)							\nonumber \\
+\frac{dq_\star}{q_\star}\frac{\zeta_q}{\phi(\zeta)}\frac{\partial \phi}{\partial \zeta}(\zeta)		,					
\end{eqnarray}
where $dT_\star$ and $dq_\star$ have been expanded in \inlinecite{ANDREAS1989} as 

\begin{eqnarray}
\frac{dT_\star}{T_\star}=S_z \frac{dz}{z}+S_{u_\star}\frac{du_\star}{u_\star}+S_{TC_{n_1}}\frac{dC_{n_1}}{C_{n_1}}+S_{TC_{n_2}}\frac{dC_{n_2}}{C_{n_2}}		,	\label{ANDREAS1989EQ1} \\
\frac{dq_\star}{q_\star}=S_z \frac{dz}{z}+S_{u_\star}\frac{du_\star}{u_\star}+S_{QC_{n_1}}\frac{dC_{n_1}}{C_{n_1}}+S_{QC_{n_2}}\frac{dC_{n_2}}{C_{n_2}}		.	\label{ANDREAS1989EQ2} 
\end{eqnarray}
Thus we have

\begin{eqnarray}
\frac{d_\epsilon}{\epsilon}=\frac{du_\star}{u_\star}\left(3+\frac{\zeta}{\phi(\zeta)}\frac{\partial \phi}{\partial \zeta}(\zeta)(S_{u_\star}-2)\right)		\nonumber \\
+\frac{dz}{z}\left(-1+\frac{\zeta}{\phi(\zeta)}\frac{\partial \phi}{\partial \zeta}(\zeta)(S_z+1)\right)+(...)\frac{dC_{n_1}}{C_{n_1}}+(...)\frac{dC_{n_2}}{C_{n_2}} ,	\label{FIFY2}
\end{eqnarray}
which gives us

\begin{eqnarray}
S_{u_\star,\epsilon}=\frac{(1/3)}{(1+\frac{1}{3}\frac{\zeta}{\phi(\zeta)}\frac{\partial \phi}{\partial \zeta}(\zeta)(S_{u_\star}-2))} 	,\label{fantastic} \\
S_{u_\star,z} = \frac{\frac{1}{3}(1-\frac{\zeta}{\phi(\zeta)}\frac{\partial \phi}{\partial \zeta}(\zeta)(S_z+1))}{(1+\frac{1}{3}\frac{\zeta}{\phi(\zeta)}\frac{\partial \phi}{\partial \zeta}(\zeta)(S_{u_\star}-2))} 	,	\label{fantastic2}
\end{eqnarray}
where the terms $(S_{u_\star}-2)$ and $(S_z+1)$ are $(S_{u_\star}-4)$ and $(S_z+2)$ in \inlinecite{ANDREAS1992}.
Eqs. \ref{fantastic} and \ref{fantastic2} reduce to Eqs. \ref{Sustarepsilon} and \ref{Sustarz}.
Also from \inlinecite{ANDREAS1989} we have

\begin{eqnarray}
S_{u_\star}=\frac{2b\zeta}{3-2b\zeta}		,	\label{sandreas1}	\\
S_z=\frac{1-2b\zeta}{3-2b\zeta}		,	\label{sandreas2}	
\end{eqnarray}
where $S_{u_\star}$ would be denoted here as $S_{T_\star,u_\star}$ and $S_z$ would be written here as $S_{T_\star,z}$ for a large-aperture scintillometer strategy not
involving the derivation of $u_\star$ from Eq. \ref{ep}. Eqs. \ref{sandreas1} and \ref{sandreas2} can be derived directly from the expressions 
in \inlinecite{ANDREAS1989} or they can be derived using the methodology outlined in this study. An alternative to using the results from \inlinecite{ANDREAS1989}
in Eqs. \ref{ANDREAS1989EQ1} and \ref{ANDREAS1989EQ2} is to 
perform the total-differential expansion in \inlinecite{ANDREAS1992} from all the equations including an expansion of Eqs. \ref{te} and \ref{qe}, although the results are the same as here.
\next

%%%%%%%%%%%%%%%%%%%%%%%%%%%%%%%%%%%%%%%%%%%%%%%%%%%%%%%%%%%%%%%%%%%%%%%%%%%%

% The endnotes section will be placed here.

%\theendnotes

\end{article}

\begin{thebibliography}{}

\bibitem[\protect\citeauthoryear{Agarwal et al.}{2001}]{FIXEDPOINT}
Agarwal RP,Meehan M,O'Regan D (2001)
\newblock {Fixed Point Theory and Applications}.
\newblock { Cambridge University Press, Cambridge}, 184 pp

\bibitem[\protect\citeauthoryear{Andreas}{1988}]{ANDREAS1988}
Andreas EL (1988)
\newblock {Estimating $C_n^2$ over snow and sea ice from meteorological data}.
\newblock { J Opt Soc Amer A} 5:481--495.

\bibitem[\protect\citeauthoryear{Andreas}{1989}]{ANDREAS1989}
Andreas EL (1989)
\newblock {Two-Wavelength Method of Measuring Path-Averaged Turbulent Surface Heat Fluxes}.
\newblock { J Atmos Oceanic Tech} 6:280--292.

\bibitem[\protect\citeauthoryear{Andreas}{1990}]{ANDREAS3}
Andreas EL (1990)
\newblock {Three-Wavelength Method of Measuring Path-Averaged Turbulent Heat Fluxes}.
\newblock { J Atmos Oceanic Tech} 7(6):801--814

\bibitem[\protect\citeauthoryear{Andreas}{1992}]{ANDREAS1992}
Andreas EL (1992)
\newblock {Uncertainty in a Path Averaged Measurement of the Friction Velocity $u_\star$}.
\newblock { J Appl Meteorol} 31:1312--1321

\bibitem[\protect\citeauthoryear{Andreas}{2012}]{ANDREAS2012}
Andreas EL (2012)
\newblock {Two Experiments on Using a Scintillometer to Infer the Surface Fluxes of Momentum and Sensible Heat}.
\newblock { J Appl Meteorol Climatol} 51:1685--1701

\bibitem[\protect\citeauthoryear{Beyrich et al.}{2002}]{BEYRICH2002}
Beyrich F, deBruin HAR, Meijninger WML, Schipper JW, Lohse H (2002)
\newblock {Results From One-Year Continuous Operation of a Large Aperture Scintillometer Over a Heterogeneous Land Surface}.
\newblock { Boundary-Layer Meteorol} 105:85--87

\bibitem[\protect\citeauthoryear{Brown and Churchill}{2009}]{COMPLEX}
Brown JW, Churchill RV (2009)
\newblock {Complex Variables and Applications, 8th Edition}.
\newblock { McGraw-Hill Book Company, New York, New York}, 468 pp

\bibitem[\protect\citeauthoryear{deBruin et al.}{2002}]{DEBRUIN2002}
deBruin HAR, Meijninger WML, Smedman A-S, Magnusson M (2002)
\newblock {Dispaced-Beam Small Aperture Scintillometer Test. Part I: The WINTEX Data-Set}.
\newblock { Boundary-Layer Meteorol} 105:129--148

\bibitem[\protect\citeauthoryear{Edwards}{1984}]{GALOIS}
Edwards HM (1984)
\newblock {Galois Theory}.
\newblock { Springer Graduate Texts in Mathematics}, 185 pp

\bibitem[\protect\citeauthoryear{Foken}{2010}]{FOKEN}
Foken T, Mauder M, Liebethal C, Wimmer F, Beyrich F, Leps J-P, Raasch S, deBruin HAR, Meijninger WML, Bange J (2010)
\newblock {Energy Balance Closure for the LITFASS-2003 Experiment}.
\newblock { Theor Appl Climatol} 101:149--160

\bibitem[\protect\citeauthoryear{Hartogensis et al.}{2002}]{HARTOGENSIS}
Hartogensis OK, deBruin HAR, Van de Wiel BJH (2002)
\newblock {Dispaced-Beam Small Aperture Scintillometer Test. Part II: CASES-99 Stable Boundary-Layer Experiment}.
\newblock { Boundary-Layer Meteorol} 105:149--176

\bibitem[\protect\citeauthoryear{Hartogensis et al.}{2003}]{HARTOGENSIS2003}
Hartogensis OK, Watts CJ, Rodriguez J-C, deBruin HAR (2003)
\newblock {Derivation of an Effective Height for Scintillometers: La Poza Experiment in Northwest Mexico}.
\newblock { J Hydrometeorol} 4:915--928

\bibitem[\protect\citeauthoryear{Hill}{1982}]{HILL1982SPACIAL}
Hill RJ (1982)
\newblock {Theory of Measuring the Path-Averaged Inner Scale of Turbulence by Spatial Filtering of Optical Scintillation}.
\newblock { Appl Optics} 21(7):1201--1211

\bibitem[\protect\citeauthoryear{Hill}{1988}]{HILL1988L0}
Hill RJ (1988)
\newblock {Comparison of Scintillation Methods for Measuring the Inner Scale of Turbulence}.
\newblock { Appl Optics} 27(11):2187--2193

\bibitem[\protect\citeauthoryear{Hill}{1989}]{HILL1989}
Hill RJ (1989)
\newblock {Implications of Monin-Obukhov similarity theory for scalar quantities}.
\newblock { J Atmos Sci} 46:2236--2244

\bibitem[\protect\citeauthoryear{Hill et al.}{1992}]{HILLSINGLEEQ}
Hill RJ, Ochs GR, Wilson JJ (1992)
\newblock {Heat and Momentum Using Optical Scintillation}.
\newblock { Boundary-Layer Meteorol} 58:391--408

\bibitem[\protect\citeauthoryear{Hoedjes et al.}{2002}]{HOEDJESOLIVES}
Hoedjes JCB, Zuurbier RM, Watts CJ (2002)
\newblock {Large Aperture Scintillometer Used over a Homogeneous Irrigated Area, Partly Affected by Regional Advection}.
\newblock { Boundary-Layer Meteorol} 105:99--117

\bibitem[\protect\citeauthoryear{Kleissl et al.}{2008}]{KLEISSL2008}
Kleissl J, Gomez J, Hong S-H, Hendrickx JMH, Rahn T, Defoor WL (2008)
\newblock {Large Aperture Scintillometer Intercomparison Study}.
\newblock { Boundary-Layer Meteorol} 128:133--150

\bibitem[\protect\citeauthoryear{Lagouarde et al.}{2002}]{LAGOUARDE}
Lagouarde JP, Bonnefond JM, Kerr YH, McAneney KJ, Irvine M (2002)
\newblock {Integrated Sensible Heat Flux Measurements of a Two-Surface Composite Landscape Using Scintillometry}.
\newblock { Boundary-Layer Meteorol} 105:5--35

\bibitem[\protect\citeauthoryear{Liu et al.}{2011}]{LIU}
Liu SM, Xu ZW, Wang WZ, Jia ZZ, Zhu MJ, Bai J, Wang JM (2011)
\newblock {A Comparison of Eddy-Covariance and Large Aperture Scintillometer Measurements With Respect to the Energy Balance Closure Problem}.
\newblock { Hydrol Earth Syst Sci} 15:1291--1306

\bibitem[\protect\citeauthoryear{Marx et al.}{2008}]{MARX}
Marx A, Kunstmann H, Sch\"{u}ttemeyer D, Moene AF (2008)
\newblock {Uncertainty Analysis for Satellite Derived Sensible Heat Fluxes and Scintillometer Measurements Over Savannah Environment and Comparison to Mesoscale Meteorological 
Simulation Results}.
\newblock { Agric For Meteorol} 148:656--667

\bibitem[\protect\citeauthoryear{Mason}{1987}]{MASON1987}
Mason, PJ (1987)
\newblock {The Formation of Areally-Averaged Roughness Lengths}.
\newblock { Q J R Meteorol Soc} 114:399-420

\bibitem[\protect\citeauthoryear{Meijninger et al.}{2002}]{MEIJNINGERPWF2002}
Meijninger WML, Hartogensis OK, Kohsiek W, Hoedjes JCB, Zuurbier RM, deBruin HAR (2002)
\newblock {Determination of Area-Averaged Sensible Heat Fluxes With A Large Aperture Scintillometer Over a Heterogeneous Surface - Flevoland Field Experiment}.
\newblock { Boundary-Layer Meteorol} 105:37--62

\bibitem[\protect\citeauthoryear{Moroni et al.}{1990}]{MORONI}
Moroni C, Navarra A, Guzzi R (1990)
\newblock {Estimation of the Turbulent Fluxes in the Surface Layer Using the Inertial Dissipative Method: a Monte Carlo Error Analysis}.
\newblock { Appl Optics} 6:2187--2193

\bibitem[\protect\citeauthoryear{Panofsky and Dutton}{1984}]{PANOFSKY}
Panofsky HA, Dutton JA, (1984)
\newblock {Atmospheric Turbulence: Models and Methods for Engineering Applications}.
\newblock {J. Wiley, New York, New York}, 397 pp

\bibitem[\protect\citeauthoryear{Press et al.}{1992}]{PRESSNUM}
Press WH, Teukolsky SA, Vetterling WT, Flannery BP (1992)
\newblock {Numerical Recipes in Fortran: The Art of Scientific Computing, 2nd Edition}.
\newblock { Cambridge University Press, Cambridge}, 963 pp

\bibitem[\protect\citeauthoryear{Sasiela}{1994}]{SASIELA}
Sasiela RJ (1994)
\newblock {Electromagnetic Wave Propagation in Turbulence: Evaluation and Application of Mellin Transforms}.
\newblock { Springer-Verlag}, 300 pp

\bibitem[\protect\citeauthoryear{Sokolnikoff}{1939}]{CHAINRULE}
Sokolnikoff IS (1939)
\newblock {Advanced Calculus}.
\newblock {McGraw-Hill Book Company, New York, New York}, 446 pp

\bibitem[\protect\citeauthoryear{Solignac et al.}{2009}]{SOLIGNAC}
Solignac PA, Brut A, Selves J-L,  B\'eteille J-P, Gastellu-Etchegorry J-P, Keravec P, B\'eziat P, Ceschia E (2009)
\newblock {Uncertainty Analysis of Computational Methods for Deriving Sensible Heat Flux Values from Scintillometer Measurements}.
\newblock { Atmos Meas Tech } 2:741--753

\bibitem[\protect\citeauthoryear{Sorbjan}{1989}]{SORBJAN}
Sorbjan Z (1989)
\newblock {Structure of the Atmospheric Boundary Layer}.
\newblock { Prentice-Hall, Englewood Cliffs, New Jersey}, 317 pp

\bibitem[\protect\citeauthoryear{Tatarski}{1961}]{TATARSKI}
Tatarski VI (1961)
\newblock {Wave Propagation in a Turbulent Medium}.
\newblock { McGraw-Hill Book Company, New York, New York}, 285 pp

\bibitem[\protect\citeauthoryear{Taylor}{1997}]{TAYLOR}
Taylor J (1997)
\newblock {An Introduction to Error Analysis: The Study of Uncertainties in Physical Measurements, 2nd edition}.
\newblock { University Science Books, Sausalito, California}, 327 pp

\bibitem[\protect\citeauthoryear{Traub}{1964}]{ITERATIVE}
Traub JF (1964)
\newblock {Iterative Methods for the Solution of Equations}.
\newblock { Prentice-Hall, Englewood Cliffs, New Jersey}, 310 pp

\bibitem[\protect\citeauthoryear{Wieringa}{1986}]{WIERINGA1986}
Wieringa J, (1986)
\newblock {Roughness-Dependent Geographical Interpolation of Surface Wind Speed Averages}.
\newblock { Q J R Meteorol Soc} 112:867-889

\bibitem[\protect\citeauthoryear{Wyngaard et al.}{1971}]{WYNGARD1971}
Wyngaard JC, Izumi Y, Collins Jr. SA (1971)
\newblock {Behavior of the Refractive Index Structure Parameter Near the Ground}.
\newblock { J Opt Soc Amer} 61:1646--1650

\bibitem[\protect\citeauthoryear{Wyngaard and Cot\'e}{1971}]{WYNGARD1971phi}
Wyngaard JC, Cot\'e OR (1971)
\newblock {The Budgets of Turbulent Kinetic Energy and Temperature Variance in the Atmospheric Surface Layer}.
\newblock { J Atmos Sci} 28:190--201

\bibitem[\protect\citeauthoryear{Wyngaard and Clifford}{1978}]{WYNGARD1978}
Wyngaard JC, Clifford SF (1978)
\newblock {Estimating Momentum, Heat and Moisture Fluxes from Structure Parameters}.
\newblock { J Atmos Sci} 35:1204--1211







\end{thebibliography}
\end{document}